\documentclass[twocolumn]{aastex701}
\usepackage{amsmath}
\usepackage{gensymb}

\begin{document}

\title{Energy partition in collisionless counterstreaming plasmas}

\author[orcid=0000-0003-3184-7215]{Alexis Marret}
\affiliation{High Energy Density Science Division, SLAC National Accelerator Laboratory, Menlo Park, California 94025, USA; \href{mailto:alexis.marret@slac.stanford.edu}{alexis.marret@slac.stanford.edu}}
\email{alexis.marret@slac.stanford.edu}  

\author[orcid=0000-0002-8502-5535]{Frederico Fiuza} 
\affiliation{GAP/Instituto de Plasmas e Fusão Nuclear, Instituto Superior Técnico, Universidade de Lisboa, 1049-001 Lisbon, Portugal; \href{mailto:frederico.fiuza@tecnico.ulisboa.pt}{frederico.fiuza@tecnico.ulisboa.pt}}
\email{frederico.fiuza@tecnico.ulisboa.pt}

\begin{abstract}

Fast, counter-streaming plasma outflows drive magnetic field amplification, plasma heating, and particle acceleration in numerous astrophysical environments, from supernova remnant shocks to active galactic nuclei jets. Understanding how, in the absence of Coulomb collisions, energy is redistributed between the different plasma species remains a fundamental open question. We use 3D fully-kinetic simulations to investigate energy partition in weakly magnetized counter-propagating plasmas. Our results reveal a complex interplay between different processes, where at early times the Weibel instability drives a first stage of magnetic field amplification and at late times the kinking of current filaments drives a second amplification stage via a dynamo-type mechanism. Electrons are heated primarily during the latter phase through magnetic pumping. By the time the flows thermalize, we observe that the final temperature ratio $T_e/T_i$ and energy partition depend on the ion-to-electron mass ratio. For electron-proton flows, the electron thermal energy only reaches up to a few percent of the initial ion kinetic energy.

\end{abstract}



\section{Introduction}

The collisionless interaction of counter-streaming flows is pervasive in astrophysical plasmas, giving rise to instabilities that shape energy partition in different systems. Observations of high-energy astrophysical environments, such as supernova remnants, colliding-wind binary systems, pulsar wind nebulae, gamma-ray bursts and active galactic nuclei jets all reveal strong non-thermal emission \citep{koyamaEvidenceShockAcceleration1995,debeckerCatalogueParticleacceleratingCollidingwind2013,h.e.s.s.collaborationDetectionVeryhighenergyGray2020,abdoSECONDFERMILARGE2013,ansoldiTeraelectronvoltPulsedEmission2016,abdallaVeryhighenergyComponentDeep2019,vieyroCollectiveNonthermalEmission2017}, magnetic field amplification \citep{berezhkoDirectEvidenceEfficient2004,reynoldsEfficienciesMagneticField2021,doughertyRadioEmissionModels2003,miaoReciprocatingMagneticFields2023,barniolduranConstrainingMagneticField2014,pudritzMagneticFieldsAstrophysical2012} and plasma heating \citep{ghavamianElectronIonTemperatureEquilibration2013,marcowithMicrophysicsCollisionlessShock2016,raymondElectronIonTemperature2023}. 

For high-Mach number (weakly magnetized) flows, fast kinetic instabilities, such as the Weibel or current-filamentation instability \citep{weibelSpontaneouslyGrowingTransverse1959,friedMechanismInstabilityTransverse1959}, are thought to control the early time flow interaction, mediating the formation of shock waves and impacting their emission properties \citep{medvedevGenerationMagneticFields1999,spitkovskyStructureRelativisticCollisionless2007}. However, despite significant progress over the last two decades, it is not yet clear what controls the thermalization of the flows and the resulting temperature ratio between electrons and ions \citep{ghavamianPhysicalRelationshipElectronProton2006,gedalinEfficientElectronHeating2008,gedalinElectronHeatingFilamentary2012,plotnikovParticleTransportHeating2013,vanthieghemOriginIntenseElectron2022,raymondElectronIonTemperature2023,vanthieghemElectronHeatingHigh2024}.

In this Letter, we investigate energy partition between ions and electrons in counter-streaming collisionless plasmas by means of three-dimensional (3D) fully kinetic particle-in-cell (PIC) simulations of weakly magnetized flows. We show that the early interaction phase is controlled by a sequence of different linear streaming instabilities that drive the amplification of electric- and magnetic-field fluctuations, albeit without significant energy exchange between plasma species. We show that the electron and ion heating results primarily from the non-linear interaction between drift-kink and current filament merging instabilities, which further amplify the magnetic field via a dynamo-type process and heat electrons via magnetic pumping. This process is intrinsically 3D and the resulting energy partition depends on the mass ratio between the plasma species. We show that the underlying processes and scalings are robust across a wide range of magnetizations.

\section{Simulation setup} 
\label{sec:style}

We performed the simulations using the OSIRIS 4.0 PIC code \citep{fonsecaOSIRISThreeDimensionalFully2002,fonsecaOnetooneDirectModeling2008}. We modeled the interaction of two symmetric, uniform, counter-streaming plasmas, each with density $n_0$, moving along the $x$ direction. We considered initial flow velocities in the range $v_0/c= [0.1 - 0.5]$, ambient magnetic fields corresponding to 
Alfv\'en Mach numbers in the range $M_A=[50-\infty]$,
and ion-to-electron mass ratios in the range $m_i/m_e = [16-1024]$.
We used 8 particles per cell per species for each flow with third-order interpolation for improved numerical accuracy.
The 3D simulation box has periodic boundary conditions and dimensions $L_x L_y L_z = 40\times 20\times 20 (c/\omega_{pi})^3$ with a spatial resolution of $0.5 c/\omega_{pe}$ in all directions, where $\omega_{pe}$ and $\omega_{pi}$ are the electron and ion plasma frequencies. Convergence was verified by repeating the simulations at a higher resolution of $0.25 c/\omega_{pe}$, which yielded consistent results.

We focus our discussion on the processes that shape the energy partition between the plasma species, using as reference the simulation with $M_A\to\infty$ (\emph{i.e.}, flows are initially unmagnetized), $m_i/m_e= 64$, and $v_0/c=0.5$ to illustrate the evolution of the system. We have verified that the sequence of plasma processes that determines the evolution and energy partition of the different species is qualitatively unchanged for the range of parameters considered. At the end, we will compare the results for the different mass ratios and Mach numbers and discuss their quantitative impact, namely on the electron-to-ion temperature ratio after the plasma flows fully thermalize.

\section{Early time evolution} 
\label{sec:floats}

\begin{figure}
    \includegraphics[width=\columnwidth]{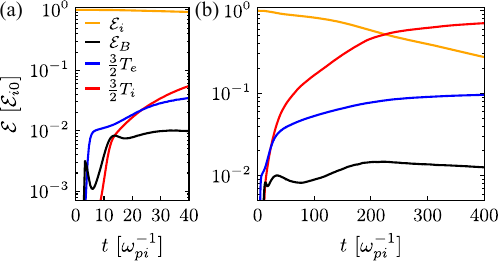}
    \caption{Evolution of the ion drift kinetic energy $\mathcal E_i=(\gamma_i-1)m_ic^2$ (orange), magnetic field energy $\mathcal E_B=B^2/(8\pi n_0)$ (black), electron temperature $T_e$ (blue) and ion temperature $T_i$ (red) for the early (a) and late (b) times, normalized to the initial ion kinetic energy $\mathcal E_{i0}$.}
    \label{fig:figure_1}
\end{figure}

As the collisionless plasmas interpenetrate, we observe the development of a sequence of streaming instabilities, consistent with linear theory and previous simulations studies \citep[e.g.,][]{achterbergWeibelInstabilityRelativistic2007a,ruyerNonlinearDynamicsIon2015}. First, the electron two-stream and Weibel 
instabilities develop on time scales of $\omega_{pe}^{-1}$ and $(c/v_0) \omega_{pe}^{-1}$ respectively, causing the amplification of electric and magnetic field fluctuations [Fig. \ref{fig:figure_1} (a), $t \approx 4\omega_{pi}^{-1}$]. During this fast phase, ions do not have time to respond. 
At saturation, we measure that the electrons convert $\approx 60\%$ of their initial drift kinetic energy $\mathcal E_{e0}$ to heat and $\approx 19\%$ to magnetic field energy $\mathcal E_B=B^2/(8\pi n_0)$. This is in good agreement with the magnetic trapping prediction $\mathcal E_B/\mathcal E_{e0}=2/\pi^2\approx20\%$ for perturbations on electron spatial scales $c/\omega_{pe}$ \citep{davidsonNonlinearDevelopmentElectromagnetic1972}. The associated fast electron bouncing motion leads to the disruption of the filaments, causing a dip in magnetic field energy at $t\approx 5\omega_{pi}^{-1}$ \citep{achterbergWeibelInstabilityRelativistic2007a}.

The ion Weibel instability then develops, producing current filaments with alternating polarity (\emph{i.e.}, current direction) and further amplifying the magnetic field energy on time scales of several $(c/v_0) \omega_{pi}^{-1}$ [Fig. \ref{fig:figure_1} (a), $t \approx 12\omega_{pi}^{-1}$]. 
The electrons move primarily along the current filaments in an effort to screen the ion current \citep{achterbergWeibelInstabilityRelativistic2007a}. For unmagnetized electrons, the resulting total current flowing along the filaments can be expressed as 
\begin{equation}
J=en_0\kappa v_i,
\label{eq:screened_current}
\end{equation}
where $\kappa=2I_1(R\omega_{pe}/c)K_1(R\omega_{pe}/c)\approx c/(R\omega_{pe})$ for $R>c/\omega_{pe}$ with $R$ the filament radius and $I_1$, $K_1$ are the modified Bessel functions of the first and second kind of order one. Here, and in the following, we refer to the electrons being unmagnetized when $kr_{Le}>1$ where $k$ is the transverse wavenumber of the filaments and $r_{Le}$ the electron gyroradius. Current screening will modify the ion Weibel instability growth rate
\begin{equation}
\Gamma_{iW} = \frac{v_0}{c}\omega_{pi}\left[1+\left(\frac{\omega_{pe}}{kc}\right)^2 \right]^{-1/2},
\end{equation}
derived under the assumption $\omega/(\sqrt{2}v_{Te}k)<1$ where $v_{Te}$ is the electron thermal velocity \citep{stixWavesPlasmas1992,nishigaiMachNumberDependence2021}, for which the maximum growth rate $\Gamma_{iW}=v_0\omega_{pi}/c$ occurs at very small, electron scales $\lambda_{iW} < 2\pi c/\omega_{pe}$. In the simulations, we measure a magnetic field growth rate $\Gamma=0.2\omega_{pi}$ and wavelength $\lambda=4.8c/\omega_{pe}$, in reasonable agreement with these predictions. Similarily to the electron Weibel phase, we find that magnetic field saturation is governed by the magnetic trapping condition $\Gamma_{iW}=\omega_{Bi}$ where $\omega_{Bi}=(\pi ev_0B/(2Rm_ic))^{1/2}$ is the ion bouncing frequency in the filaments
as
\begin{equation}
B_{trap}\approx\frac{2m_icR\Gamma_{iW}^2}{\pi ev_0}.
\label{eq:Bsat_t}
\end{equation}
This condition is independent of the ion mass, preventing amplification significantly beyond the level of the electron Weibel instability. 
Given the weak magnetic field level and the small scale of the filaments, the electrons remain unmagnetized.
Both electron and ion heating remain low at this stage, \emph{i.e.,} $\approx 1 \%$ of the initial kinetic energy of the flow for $m_i/m_e=64$.
These results are confirmed by our simulations for all ion-to-electron mass ratios investigated.

After saturation of the ion Weibel instability, the plasma is organized in a network of ion current filaments partially screened by the unmagnetized electrons. The relative drift velocity between electrons and ions in the filaments triggers the Buneman instability [Fig. \ref{fig:figure_1} (a), saturation at $t \approx 30\omega_{pi}^{-1}$] that further heats the electrons and moderately amplifies the magnetic fields \citep{treumannMagneticFieldAmplification2012}.
However, it saturates quite rapidly as the electron thermal velocity becomes larger than the electron-ion relative velocity so that $T_e\approx m_ev_0^2\ll m_iv_0^2$, not being able to extract a significant fraction of the ion kinetic energy.

\section{Late time evolution} 
\label{sec:highlight}

It is during the nonlinear evolution of the ion current filaments that most of the electron and ion heating will occur [Fig. \ref{fig:figure_1} (b)]. Let us then discuss what controls the evolution of the system in this phase. Filaments with same polarity start attracting each other via the magnetic force, eventually leading to merging and to an inverse cascade \citep{medvedevLongTimeEvolutionMagnetic2005,achterbergWeibelInstabilityRelativistic2007,ruyerNonlinearDynamicsIon2015}. In addition, the non-linear filament dynamics is significantly modified by the growth of the drift-kink instability, which produces $m=1$ transverse helical perturbations of the current filaments \citep{ruyerDisruptionCurrentFilaments2018,vanthieghemStabilityAnalysisPeriodic2018}. As we will discuss below, and contrary to the previous understanding, we find that this late phase governed by the kink mode is central to both magnetic field amplification and energy partition.

\begin{figure}
    \includegraphics[width=\columnwidth]{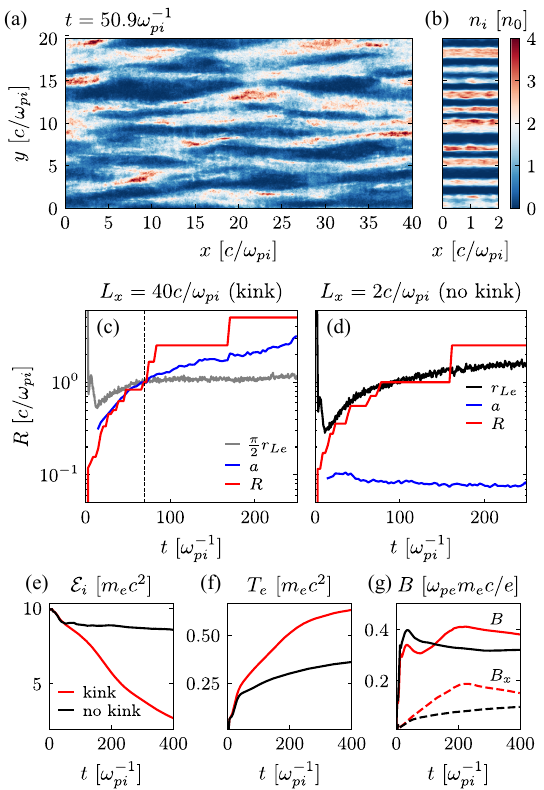}
    \caption{Ion density ($x-y$) slice for the ion population propagating initially toward the positive $x$ direction in the (a) full box and (b) shortened box cases, taken at $t=50.9\omega_{pi}^{-1}$ during the drift-kink instability phase. (c) Temporal evolution of filament radius $R$ (red) and comparison with the size of the transverse perturbation amplitude of the filaments $a$ (blue) and with the scaled electron gyroradius extracted from particle tracking data $r_{Le}(\pi/2)$ (gray) for the full box case. The dashed vertical line indicates the time when the electrons become magnetized $r_{Le}(\pi/2)\leq R$. (d) Similar to (c) but for the shortened box case $L_x=2c/\omega_{pi}<\lambda_{kink}$. The black line indicates the electron gyroradius $r_{Le}$.  (e) Evolution of the ion drift kinetic energy $\mathcal E_i$, (f) electron temperature $T_e$, (g) total magnetic field $B$ (solid line) and longitudinal component $B_x$ (dashed line), in the full box (red line, kink) and shortened box (black line, no kink) cases.} 
    \label{fig:figure_2}
\end{figure}

The development of the drift-kink instability is confirmed in the simulations with the observation of transverse modulations of the filaments with a longitudinal wavelength  $\lambda\approx 15c/\omega_{pi}$ [Fig. \ref{fig:figure_2} (a)]. This is in good agreement with the resonant wavelength $\lambda_{kink}=2\pi v_0/\omega_{Bi}$, predicted from linear kinetic theory \citep{daughtonKineticTheoryDrift1998,ruyerDisruptionCurrentFilaments2018} for the magnetic field amplitude and filament radius measured in the simulation [Fig. \ref{fig:figure_2} (c)]. We note that this is much larger than the fluid prediction of $\lambda_{kink} \approx R$ \citep{daughtonTwofluidTheoryDrift1999}.

Interestingly, these transverse kink perturbations accelerate the interaction and merging between filaments, which in turn constrain the amplitude of the perturbations. 
In the case of an isolated filament, the drift-kink instability will quickly disrupt the filament when the transverse displacement amplitude $a$ exceeds the radius $R$ \citep{ruyerDisruptionCurrentFilaments2018}. However, for multiple filaments, such a large displacement will lead to a collision and a merging with adjacent filaments, thus increasing $R$ and maintaining the condition $a\leq R$. This is confirmed in Figure \ref{fig:figure_2} (c) where we observe the increase of the filament radius in the simulation over time up to $R>2c/\omega_{pi}$ ($t>100\omega_{pi}^{-1}$), while the perturbations amplitude verifies $a\leq R$ at all times. 
This helps explain how the filaments can be sustained on times scales of $100s\ \omega_{pi}^{-1}$, significantly longer than the resonant drift-kink instability growth time \citep{ruyerDisruptionCurrentFilaments2018}
\begin{equation}
\Gamma_{kink}^{-1}=\left(\frac{3}{2}\frac{v_0}{c}\sqrt{\frac{m_e}{m_i}\frac{c}{R\omega_{pe}}}\omega_{pi}\right)^{-1},
\end{equation}
which yields $\Gamma_{kink}^{-1}\approx 20-30\omega_{pi}^{-1}$ for $m_i/m_e=64$ and for the range of filament radii extracted from the simulation. [Here, we measured the displacement $a$ by identifying the centroids of the filaments in a large number of $(x-y)$ slices of the simulation domain, and by computing for each filament the largest departure from its mean $y$ position which is then averaged over all identified filaments.]

To further clarify the importance of the drift-kink instability on the thermalization of the flows, we disable the growth of the kink perturbations by performing additional 3D simulations with all parameters the same except for a shortened longitudinal box size $L_x=2c/\omega_{pi}< \lambda_{kink}$ [Fig. \ref{fig:figure_2} (b)], which prevents the instability development. In this case, most of the ion drift kinetic energy is conserved for the duration of the simulation and both ion and electron heating is largely reduced [Fig. \ref{fig:figure_2} (e) and (f)]. The current filaments remain straight ($a\ll R$) and the merging rate decreases rapidly after a few merging steps [Fig. \ref{fig:figure_2} (d)] \citep{medvedevLongTimeEvolutionMagnetic2005,zhouMultiscaleDynamicsMagnetic2020}.

In the absence of kink perturbations, the slow electron heating after saturation of the early stages instabilities [Fig. \ref{fig:figure_2} (f), $t>30\omega_{pi}^{-1}$] is due to the chaotic electron motion in the potential associated with the filament $A_x\approx RB_{z}$, which leads to equipartition between the kinetic and potential energy under the approximate conservation of the canonical momentum $P_x=m_ev_{x}-e A_x/c$ \citep{bresciSaturationAsymmetricCurrent2022}. This corresponds to $r_{Le}\approx R$, in good agreement with the simulations [Fig. \ref{fig:figure_2} (d)]. 
The electrons remain marginally unmagnetized at all times ($r_{Le}>2R/\pi$) 
and can screen the current such that $J\propto 1/R$ [Eq. \ref{eq:screened_current}]. The magnetic field associated to the filaments 
$B=2\pi JR/c$ 
is then expected to be constant during merging, in agreement with the shortened box simulations [Fig. \ref{fig:figure_2} (g)] and in contrast with the full box case where an additional phase of magnetic field amplification occurs.

The approximate conservation of canonical momentum is valid under the assumption that the electron Lagrangian is independent of the position along the filament and no longer holds if the filaments are perturbed by the drift-kink instability. In the full box simulation (where kink perturbations can develop) the electrons can reach the condition $r_{Le}<2R/\pi$, and become magnetized [Fig. \ref{fig:figure_2} (c), $t \approx 70\omega_{pi}^{-1}$]. 
In addition to stopping the current screening \citep{nishigaiMachNumberDependence2021}, we find that the electron magnetization in the kinking filaments enables the late time magnetic field amplification and significant electron heating.

\section{Electron heating and magnetic field amplification}

The electron heating during the late time evolution coincides with a magnetic field amplification stage, that further increases the fields produced by the early time instabilities [Fig. 2 (f) and (g), $t>70\omega_{pi}^{-1}$]. The magnetic field and electron temperature are observed to be proportional during this phase and follow, as we will show later, the same scaling $B\propto T_e\propto m_i^{1/2}$ at the end of the simulations. 
These results point toward betatron acceleration via magnetic pumping as being the dominant mechanism for heating the magnetized electrons during the late stages of the plasma evolution. 
Assuming negligible density perturbations within the filaments, the resulting electron heating rate can be calculated if the magnetic field amplification rate is known following the Chew-Goldberger-Low equation \citep{chewBoltzmannEquationOnefluid1956,vaskoElectronHolesInhomogeneous2016}
\begin{equation}
\frac{d T_{e}}{d t} = \frac{T_{e0}}{B_{0}}\frac{dB}{dt},
\label{eq:dTedt_theory}
\end{equation}
where $T_{e0}$ and $B_{0}$ are the temperature and magnetic field amplitude when the electrons become magnetized. We note that magnetic pumping heats electrons primarily in the direction perpendicular to the magnetic field, but we verified that electrons quickly isotropize their temperature by scattering in the filaments such that $T_{e\perp}\approx T_e$.

\begin{figure}
    \includegraphics[width=\columnwidth]{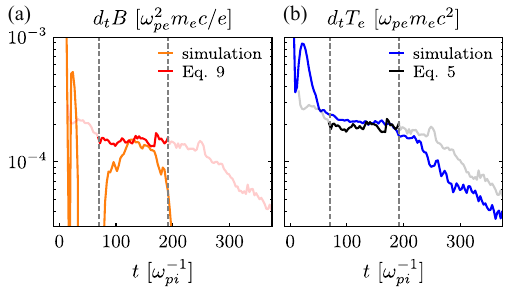}
    \caption{(a) Evolution of the magnetic field amplification rate $dB/dt$ from the simulation (orange) and from Eq. \ref{eq:dBdt_theory} (red). (b) Evolution of the electron heating rate $d T_e/d t$ from the simulation (blue) and from Eq. \ref{eq:dTedt_theory} (black). The dashed vertical lines indicate the electron magnetization ($r_{Le}(\pi/2)\leq R$, $t\approx70\omega_{pi}^{-1}$) and drift-kink instability saturation (Eq. \ref{eq:bsat_v}, $t\approx190\omega_{pi}^{-1}$), with the theoretical predictions valid only within this time window and shown with reduced opacity beyond it. 
    } 
    \label{fig:figure_3}
\end{figure}

We have found that the magnetic field amplification results from a magnetic dynamo driven by the kink transverse perturbations, that stretches the magnetic field lines associated with the filaments. To investigate this phenomenon, we develop a model consisting of a single filament perturbed by an helical ($m=1$) type kink deformation along the $x$ direction, parametrized by the perturbation amplitude $a$ and wavelength $\lambda_{kink} = 2\pi/k$. The net magnetic field amplification resulting from this geometry can be calculated with the induction equation accounting for magnetic field advection
\begin{equation}
\frac{d\mathbf B}{dt} = \nabla\times\mathbf (\mathbf v_e\times\mathbf B)+\mathbf (\mathbf v_e\cdot\nabla)\mathbf B=(\mathbf B\cdot\nabla)\mathbf v_e - \mathbf B(\nabla\cdot\mathbf v_e).
\label{eq:idealOhm_faraday}
\end{equation}
During this stage electrons are magnetized while ions remain unmagnetized. Thus the magnetic field is frozen-in to the electron fluid and the motional electric field is expressed as $\mathbf{E=-\mathbf v_e\times\mathbf B}/c$.
The electron fluid velocity $\mathbf v_e$ is aligned with the perturbed filament axis, and the magnetic field $\mathbf B$ is purely toroidal in the plane transverse to the local filament axis direction. After projecting in the laboratory frame one obtains
\begin{align}
\mathbf v_e &=\frac{v_e}{\alpha}(\mathbf e_x+C\mathbf e_y-S\mathbf e_z), \label{eq:ue_labframe}
\\
\mathbf B &=\frac{B}{\alpha R}[(y'S+z'C)\mathbf e_x-z'\mathbf e_y+y'\mathbf e_z], \label{eq:B_labframe}
\end{align}
where $v_e$ and $B$ are amplitudes, $S=ak\sin(kx)$, $C=ak\cos(kx)$, $\alpha =\sqrt{1+a^2k^2}$, and $y'=y-S/k$ and $z'=z-C/k$ are the transverse coordinates relative to the filament axis (see appendix \ref{app:dynamo_model} for details).
Since the flow is purely rotational, one finds $\nabla\cdot\mathbf v_e = 0$ and only the magnetic field stretching term $\mathbf (\mathbf B\cdot\nabla)\mathbf v_e$ contributes to the magnetic field amplification. This term is zero for a straight filament. It is the 3D kink-type deformations that produce longitudinal gradients of the flow velocity aligned with the magnetic field $B_x\partial_x(v_y\mathbf e_y+v_z\mathbf e_z)$, driving a dynamo-type amplification within the filament. 
By inserting Eqs. \ref{eq:ue_labframe} and \ref{eq:B_labframe} in Eq. \ref{eq:idealOhm_faraday}, one obtains, after some algebra, that the mean absolute magnetic amplification is
\begin{equation}
\frac{d B_y}{dt}=\frac{d B_z}{dt}=\sqrt{\frac{2}{3\pi^2}} v_e  B\frac{a^2k^2}{1+a^2k^2}k.
\label{eq:dBdt_theory}
\end{equation}
We computed Eqs. \ref{eq:dTedt_theory} and \ref{eq:dBdt_theory} using the magnetic field, electron fluid velocity and transverse displacement of the filaments extracted from the simulation, and compared it with the electron heating and magnetic field amplification observed in the simulations. This model can reproduce quantitatively the rates observed during the drift-kink phase [Fig. \ref{fig:figure_3} (a) and (b), $t\approx70-190\omega_{pi}^{-1}$]. 

While it is the transverse magnetic field component in the laboratory frame that gets amplified, due to the tilting of the filaments by the kink perturbations, a longitudinal ($B_x$) field component will naturally develop, as can be seen in Eq. \ref{eq:B_labframe} and in Fig. \ref{fig:figure_2} (g). The simulations with full box size show that $d_tB_x\propto d_t(a/\lambda_{kink})$, confirming the role of the filaments tilting in isotropizing the magnetic field, while the shortened box case exhibits much weaker $B_x$ component.

The dynamo-type mechanism described by Eq. \ref{eq:dBdt_theory} is intrinsically 3D (see appendix \ref{app:2D_simulations} for a more detailed discussion of the differences between 2D and 3D) and, to our knowledge, has not been described previously in the literature. Previous work \citep{jikeiEnhancedMagneticField2024} identified a different type of dynamo amplification, driven by the velocity shear between straight current filaments in 2D simulations of magnetized counter-streaming flows. We could not observe this effect in our simulations, most likely due to the kink perturbations that create strong inhomogeneities along the filaments axis in the full box case, and because the electrons are kept unmagnetized due to the conservation of canonical momentum in the shortened box case.

The combination of the increasing filament radius and of the magnetic field amplification gradually increases the ion bouncing velocity in the filaments. The ions are effectively scattered and the current filaments are disrupted when this velocity becomes comparable to the ion drift velocity. This criterion, corresponding to the Alfv\'en current limit \citep{achterbergWeibelInstabilityRelativistic2007a}, can be expressed as
\begin{equation}
R\omega_{Bi}\approx v_i,
\label{eq:saturation_criterion}
\end{equation}
and has been verified against the simulations for all mass ratios investigated [Fig. \ref{fig:figure_4} (a)]. The resulting maximum magnetic field amplification is
\begin{equation}
B_{A}\approx\frac{2m_icv_i}{\pi eR}.
\label{eq:bsat_v}
\end{equation}
The disruption of the filaments rapidly thermalizes the remaining ion drift kinetic energy, halts the magnetic field dynamo-type amplification [Fig. \ref{fig:figure_3} (a), $t\approx190\omega_{pi}^{-1}$], and dictates the final electron-to-ion temperature ratio, $T_e/T_i$.

By inserting in Eq. \ref{eq:bsat_v} the scaling $R\approx c/\omega_{pi}$ as observed in the simulations at the end of the drift-kink phase, one obtains $B_{A}\propto m_i^{1/2}$, and the ratio of magnetic to ion kinetic energy $\sigma_i=\mathcal E_B/\mathcal E_{i0}$ is independent of the mass ratio. For initially screened filaments, this corresponds to $B_{A}/B_{trap}\propto (m_i/m_e)^{1/2}$, so that the magnetic field produced via the dynamo-type mechanism exceeds that obtained from the screened ion Weibel instability.
If, however, the initial ambient magnetic field was to be strong enough to magnetize the electrons (see detailed discussion below), they would be unable to screen the ion current and the filament size at saturation of the ion Weibel instability would be $R\approx c/\omega_{pi}$.
In that case, one recovers $B_{A}=B_{trap}$ which is the expected result for unscreened filaments.

\section{Mass ratio scaling}

\label{sec:mass_ratio_scaling}
\begin{figure}
\includegraphics[width=\columnwidth]{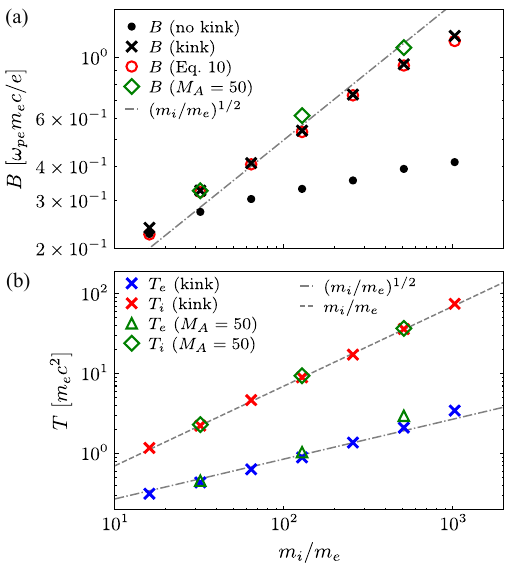}
    \caption{(a) Magnetic field amplification $B$ as a function of mass ratio $m_i/m_e$ in the shortened box (noted 'no kink', black dot markers) and full box (noted 'kink', black cross markers) cases. The magnetic field observed in the full box simulations at the time Eq. \ref{eq:bsat_v} is verified is shown with red circles. The case $M_A=50$ is shown with green square markers. (b) Electron temperature (blue) and ion temperature (red) in the full box case (noted 'kink', cross markers). The results for the case $M_A=50$ are shown with green triangles (electrons) and squares (ions).}
    \label{fig:figure_4}
\end{figure}

We have varied the ion-to-electron mass ratio used in the simulations to study how it impacts the dominant processes that shape energy partition in counter-streaming plasma flows. Figure \ref{fig:figure_4} displays the scaling with $m_i/m_e$ of the saturated magnetic field, electron temperature, and ion temperature extracted at the end of the simulations after the thermalization of the flows. 
We observe that the saturated magnetic field associated with the drift-kink instability and corresponding dynamo-type amplification scales as $(m_i/m_e)^{1/2}$, consistent with Eq. \ref{eq:bsat_v} for $R\approx c/\omega_{pi}$, and is significantly larger than that produced without the drift-kink instability for large mass ratios [Fig. \ref{fig:figure_4} (a)]. The electron temperature follows the same scaling $T_e \propto B \propto (m_i/m_e)^{1/2}$, consistent with magnetic pumping, while the ion temperature scales with the ion mass as expected $T_i \propto m_i/m_e$. The temperature ratio obtained from the simulations after full thermalization of the flows is $T_e/T_i \approx1.2 (m_i/m_e)^{-1/2}$. 
This result is compatible with previously reported values $T_e/T_i \approx [0.2-0.3]$ in numerical studies of high-Mach number counter-streaming flows where typical mass ratios $m_i/m_e \approx [16-100]$ are considered \citep{spitkovskyStructureRelativisticCollisionless2007,kumarElectronHeatingRelativistic2015,vanthieghemElectronHeatingHigh2024}. For realistic electron-proton plasmas, however, this leads to a significantly smaller value $T_e/T_i = 0.03$ meaning that electrons only receive $\approx2\%$ of the initial kinetic energy of the flows.

\section{Role of ambient magnetic field}

As highlighted previously, the electron magnetization plays a central role in controlling both the early and late stages of magnetic field amplification and electron heating. We may expect electrons to be magnetized and unable to screen the early time electron scale filaments if 
$r_{Le}<(2/\pi)c/\omega_{pe}$. This condition can be written in terms of the initial Alfv\'en Mach number of the flows $M_A$ and the angle of ambient magnetic field orientation with respect to the flow velocity $\theta_B$, which yields $M_A<(2/\pi)(m_i/m_e)^{1/2}\sin(\theta_B)$. 

To investigate the role of an ambient magnetic field, we have performed a series of 3D simulations with $m_i/m_e=[32,128,512]$ where we introduce an oblique ($\theta_B=45\degree$) magnetic field in the $(x-y)$ plane with $M_A=50$. We find that the ambient magnetic field is insufficient to magnetize the electrons during the ion Weibel growth phase, consistent with the condition $k r_{Le}<1$ for magnetization. The plasma evolution proceeds similarly to the case $M_A\to\infty$ (see appendix \ref{app:ambient_field}), and the final energy partition is not significantly modified [Fig. \ref{fig:figure_4}]. 
We note that previous work \citep{nishigaiMachNumberDependence2021} suggested a different magnetization threshold for disabling electron screening and modifying the scale of the ion Weibel filaments during the linear phase as $\Gamma_{iW}/\Omega_{ce}< 1$, where $\Omega_{ce}$ is the electron gyrofrequency, corresponding to $M_A<(m_i/m_e)\sin(\theta_B)$. This would be met by our $M_A = 50$ simulations for $m_i/m_e=[128,512]$. Our results thus favor the less stringent criterion $kr_{Le}<1$, and we expect that the results presented in this Letter will remain valid for Mach numbers $M_A>20$ for realistic mass ratios $m_i/m_e = 1836$ and $\theta_B=45\degree$. We defer to future work a comprehensive study of the energy partition dependency on Mach number and magnetic field obliquity.

\section{Conclusion}

In summary, we have shown that while the early time interaction of fast, collisionless counter-propagating plasma flows is controlled by the filamentary Weibel instability, significant electron and ion heating only occur at a later phase and are largely controlled by the growth of the drift-kink instability and an associated dynamo-type mechanism. The electrons are heated primarily by betatron acceleration which is driven by the stretching of the magnetic field lines in the kinking filaments. This mechanism is intrinsically 3D, and is robust for a wide range of initial magnetizations.

The final temperature ratio $T_e/T_i$, obtained after the flows fully thermalize, is shown to scale with $(m_i/m_e)^{-1/2}$.
The inferred temperature ratio $T_e/T_i=0.03$ for $m_i/m_e=1836$ is compatible with astrophysical observations of supernova remnants \citep{ghavamianPhysicalRelationshipElectronProton2006,ghavamianElectronIonTemperatureEquilibration2013,raymondElectronIonTemperature2023}. 
These results are also particularly timely in the context of emerging experimental platforms dedicated to studying energy partition in counter-streaming plasma flows \citep{foxFilamentationInstabilityCounterstreaming2013,huntingtonObservationMagneticField2015,swadlingMeasurementKineticScaleCurrent2020} and collisionless shocks in weakly magnetized laboratory plasmas \citep{fiuzaElectronAccelerationLaboratoryproduced2020}. 
At the same time, we should emphasize that periodic-box simulations do not capture all the shock physics and that kinetic simulations can only capture much reduced spatial and temporal scales when compared to observations. Further work comparing the results of periodic flows and shock simulations with laboratory experiments can help elucidate the full extent of the importance of these processes in collisionless shocks.

\begin{acknowledgments}
The authors acknowledge the support from the European Research Council (ERC-2021-CoG Grant XPACE No. 101045172) and by the US DOE
Early Career Research Program under FWP 100331. The authors acknowledge the OSIRIS Consortium, consisting of UCLA and IST (Portugal) for the use of the OSIRIS 4.0 framework. The simulations were run on Theta (ALCF) and Perlmutter (NERSC) through an ALCC award.
\end{acknowledgments}

\appendix
\twocolumngrid
\section{Magnetic field dynamo amplification model}
\label{app:dynamo_model}

We consider a helical filament due to kinking ($m=1$) perturbations with helix radius (\emph{i.e.} perturbation amplitude) $a$ and longitudinal wavelength $\lambda=2\pi/k$. The coordinates of the filament axis are $\mathbf c=x\mathbf e_x+a\sin(kx)\mathbf e_y+a\cos(kx)\mathbf e_z$ and the position relative to the filament axis is $x'=0$, $y'=y-c_y$ and $z'=z-c_z$. We seek to calculate the magnetic field amplification resulting from this geometry, accounting for the magnetic field advection as per Eq. \ref{eq:idealOhm_faraday}.
We calculate the fields associated with the filament using a local basis ($\mathbf e_1=\partial_x \mathbf c$, $\mathbf e_2 = \mathbf e_1\times\mathbf e_z$, $\mathbf e_3 = \mathbf e_1\times\mathbf e_2$). After normalization, one obtains 
\begin{align}
\mathbf e_1=&\ (\mathbf e_x+C\mathbf e_y-S\mathbf e_z)/\alpha,\\   
\mathbf e_2=&\ (C\mathbf e_x-\mathbf e_y)/\xi ,\\ 
\mathbf e_3=&-(S\mathbf e_x+SC\mathbf e_y+\xi^2\mathbf e_z)/\alpha\xi,
\end{align}
where we defined $S=ak\sin(kx)$, $C=ak\cos(kx)$, $\alpha =\sqrt{1+a^2k^2}$ and $\xi=\sqrt{1+C^2}$. 
The magnetic field and electron fluid velocity are defined as $\mathbf B =B(-r_3 \mathbf e_2 + r_2\mathbf e_3)/R$ and $\mathbf v_e =v_e\mathbf e_1$ within the filament, \emph{i.e.}, for $\sqrt{r_2^2+r_3^2}<R$ where $R$ is the filament radius, $r_2=(y'\mathbf e_y+z'\mathbf e_z)\cdot\mathbf e_2$ and $r_3=(y'\mathbf e_y+z'\mathbf e_z)\cdot\mathbf e_3$ are the relative cartesian coordinates in the plane transverse to the local filament axis direction, $v_e$ and $B$ are amplitudes. After projecting the fields in the laboratory frame, one obtains
\begin{align}
\mathbf B &=\frac{B}{\alpha R}[(y'S+z'C)\mathbf e_x-z'\mathbf e_y+y'\mathbf e_z],
\\
\mathbf v_e &=\frac{v_e}{\alpha}(\mathbf e_x+C\mathbf e_y-S\mathbf e_z).
\end{align}
One finds $\nabla\cdot\mathbf v_e = 0$ since the flow is purely rotational. The magnetic field stretching term $(\mathbf B\cdot\nabla)\mathbf v_e$ in Eq. \ref{eq:idealOhm_faraday} can be calculated as
\begin{align}
(\mathbf B\cdot\nabla)\mathbf v_e &= -k\dfrac{v_eB}{R\alpha^2}(y'S+z'C)(S\mathbf e_y+C\mathbf e_z).
\end{align}
One may then compute the mean square of the magnetic field amplification within the filament in the laboratory frame as
\begin{align}
\langle[(\mathbf B\cdot\nabla)\mathbf v_{e}]^2\rangle=\int_0^{\lambda}\int_{-R}^{R}\int_{-R}^{R}\dfrac{[(\mathbf B\cdot\nabla)\mathbf v_{e}]^2}{\lambda(2R)^2}dxdy'dz'.
\label{eq:integral_Bgradu}
\end{align} 
By integrating Eq. \ref{eq:integral_Bgradu} we obtain, after taking the square root,
\begin{equation}
\mathrm{RMS}\left(\dfrac{dB_y}{dt}\right)=\mathrm{RMS}\left(\dfrac{dB_z}{dt}\right)=\dfrac{v_eB}{\sqrt{6}}\dfrac{a^2k^2}{1+a^2k^2}k,
\end{equation}
which we relate to the mean absolute magnetic amplification by accounting for the periodic envelope $\int_0^\lambda|\sin(kx)|dx=\pi/2$ of the velocity field to obtain Eq. \ref{eq:dBdt_theory}. We note that there is no net amplification associated with the $B_x$ component.

\section{Two-dimensional simulations}
\label{app:2D_simulations}

\begin{figure}
\centering
\includegraphics[width=\columnwidth]{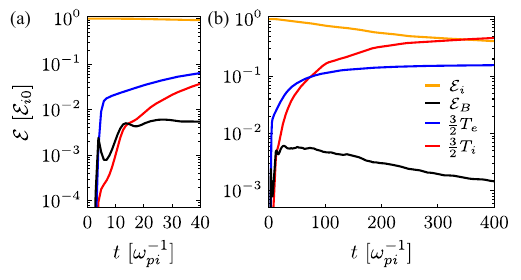}
    \caption{Evolution of the ion drift kinetic energy $\mathcal E_i$ (orange), magnetic field energy $\mathcal E_B$ (black), electron temperature $T_e$ (blue) and ion temperature $T_i$ (red) for the early (a) and late (b) times, normalized to the initial ion kinetic energy $\mathcal E_{i0}$ in a 2D simulation with $L_xL_y=40\times20(c/\omega_{pi})^2$ and $m_i/m_e=64$.}
    \label{fig:appendix_B}
\end{figure}

We investigated the importance of simulation dimensionality in modeling the thermalization of the flows by performing additional 2D in-plane simulations with dimensions $L_xL_y=40\times20(c/\omega_{pi})^2$. We considered $m_i/m_e=[16-512]$ and used $100$ particles per cell per species, while keeping all other parameters the same. The results are shown in Fig. \ref{fig:appendix_B} (a) and (b) for the case $m_i/m_e=64$. While the early time evolution, controlled by the growth of the Weibel instability, is similar to the 3D case (compare with Fig. \ref{fig:figure_1}), the late time dynamo magnetic field amplification is not recovered, confirming the importance of the 3D kinking of the filaments and associated dynamo amplification in the energy partition process. 
In 2D, electrons are primarily heated due to filament merging right after the saturation of the ion Weibel \citep{takamotoMagneticFieldSaturation2018}, and the slow down and thermalization of the ions occurs on a longer time scale than in 3D. In the end, we find that the temperature ratio $T_e/T_i$ measured is $\approx 2 \times$ that of the 3D case. A detailed description of the electron heating mechanism in the 2D case is left for future work.

\section{Simulations with ambient magnetic field}
\label{app:ambient_field}

We have performed a series of 3D simulations to study the impact of an ambient magnetic field on the flows interaction and energy partition. The simulations use the same parameters as those for initially unmagnetized plasmas but we introduce a magnetic field in the $(x-y)$ plane with an oblique angle $\theta_B=45\degree$. We simulate different ion-to-electron mass ratios $m_i/m_e=[32,128,512]$ and vary the magnetic field intensity to maintain an initial Alfv\'en Mach number $M_A=50$. 

\begin{figure}
\centering
\includegraphics[width=0.79\columnwidth]{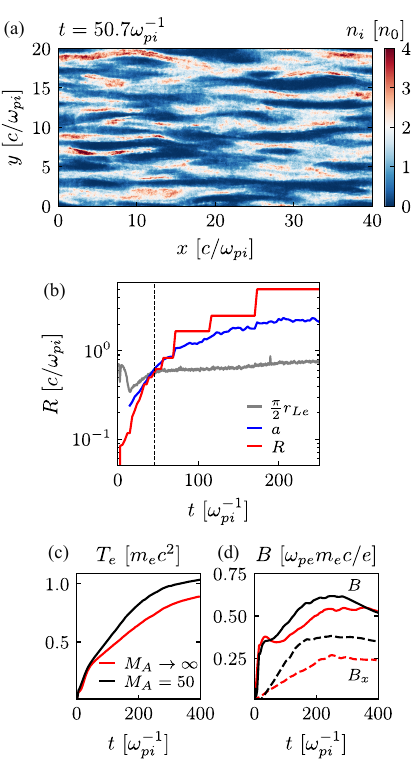}
    \caption{Ion density ($x-y$) slice for the ion population propagating initially toward the positive $x$ direction in the case $M_A=50$, taken at $t=50.7\omega_{pi}^{-1}$ during the drift-kink instability phase. (b) Temporal evolution of filament radius $R$ (red) and comparison with the size of the transverse perturbation amplitude of the filaments $a$ (blue) and with the scaled electron gyroradius $r_{Le}(\pi/2)$ (gray). The dashed vertical line indicates the time when the electrons become magnetized $r_{Le}(\pi/2)\leq R$. (c) Evolution of the electron temperature $T_e$, (d) total magnetic field $B$ (solid line) and longitudinal component $B_x$ (dashed line), in the initially unmagnetized case (red line, $M_A\to\infty$) and including an ambient magnetic field (black line, $M_A=50$). The data are taken from the simulations with $m_i/m_e=128$.}
    \label{fig:appendix_C}
\end{figure}

We find that the initial magnetic field is not sufficiently large to magnetize the electrons during the ion Weibel growth phase ($t\approx10\omega_{pi}^{-1}$), and the filaments first grow on small, electron spatial scales due to electron screening.  We also observe the development of the drift-kink instability on the same spatial scales as in the initially unmagnetized case [Fig. \ref{fig:appendix_C} (a)], leading to the fast increase of the filaments radius [Fig. \ref{fig:appendix_C} (b)] while the filament perturbation amplitude verifies $a\leq R$ at all times. The electrons become magnetized later on due to the fast increase in filaments radius ($t>45\omega_{pi}^{-1}$). We observe the late phase dynamo magnetic field amplification and associated electron heating again fully consistent with the unmagnetized case, albeit at a moderately faster rate [Fig. \ref{fig:appendix_C} (c) and (d)]. These results indicate that even for $M_A = 50$, the results discussed for the unmagnetized ($M_A \rightarrow \infty$) case hold.

\bibliography{bibliography}{}
\bibliographystyle{aasjournalv7}


\end{document}